\documentclass[conference]{IEEEtran}
\IEEEoverridecommandlockouts
\usepackage{cite}
\usepackage{amsmath,amssymb,amsfonts}
\usepackage{algorithmic}
\usepackage{graphicx}
\usepackage{textcomp}
\usepackage{xcolor}
\usepackage[T1]{fontenc}

\usepackage{caption}
\def\BibTeX{{\rm B\kern-.05em{\sc i\kern-.025em b}\kern-.08em
    T\kern-.1667em\lower.7ex\hbox{E}\kern-.125emX}}
\begin{document}

\title{Advanced Energy-Efficient System for Precision Electrodermal Activity Monitoring in Stress Detection\\

}
\author{\IEEEauthorblockN{ Ruoyu Zhang,
Ruijie Fang,
Elahe Hosseini,
Chongzhou Fang,
Ning Miao,
Houman Homayoun}
\IEEEauthorblockA{
University of California, Davis\\
Email: \{ryuzhang,rjfang,ehosseini,czfang,nmiao,hhomayoun\}@ucdavis.edu}

}




\maketitle

\begin{abstract}

This paper presents a novel Electrodermal Activity (EDA) signal acquisition system, designed to address the challenges of stress monitoring in contemporary society, where stress affects one in four individuals. Our system focuses on enhancing the accuracy and efficiency of EDA measurements, a reliable indicator of stress. Traditional EDA monitoring solutions often grapple with trade-offs between sensor placement, cost, and power consumption, leading to compromised data accuracy. Our innovative design incorporates an adaptive gain mechanism, catering to the broad dynamic range and high-resolution needs of EDA data analysis. The performance of our system was extensively tested through simulations and a custom Printed Circuit Board (PCB), achieving an error rate below 1\% and maintaining power consumption at a mere 700$\mu$A under a 3.7V power supply. This research contributes significantly to the field of wearable health technology, offering a robust and efficient solution for long-term stress monitoring.

\end{abstract}


\section{Introduction}

Nowadays, stress, as a prevalent mental health issue, impacts one in four individuals, with wide-ranging implications both mentally and physically \cite{robinson2023teachers}. Psychological conditions like depression and anxiety, and in severe cases, suicide, are linked to stress. Physically, it is associated with serious health issues, including high blood pressure, strokes, and heart attacks \cite{shrivastav2023impact}. Furthermore, research suggests stress adversely affects the immune system, potentially increasing cancer risks. Stress also significantly impacts interpersonal relationships and workplace performance, leading to reduced quality of life \cite{hammoudi2023understanding}. As a reliable stress indicator \cite{hosseini2023emotion, hosseini2022low, fang2022towards, fang2022prevent,fang2022pain},  electrodermal Activity (EDA) gaining traction in both research, and commercial product development circles, especially with the rise of wearable devices \cite{lal2023compressed,zhang2023short,zhang2021iwrap, zhang2023privee}.

\begin{figure}[ht!]
    \centering
    \includegraphics[width= \linewidth]{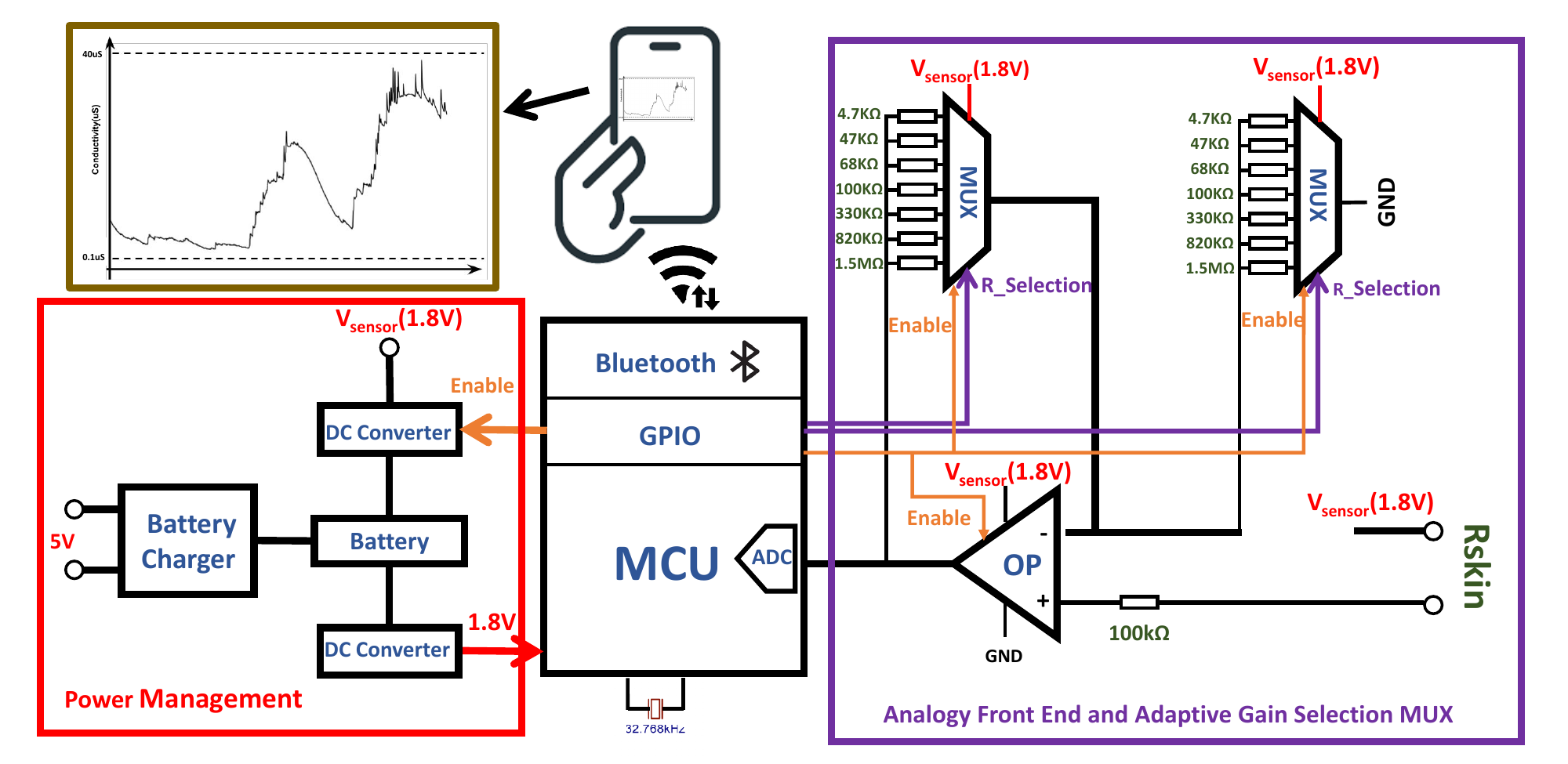}
    \caption{ Illustration of the External Circuits Modules and MCU for Electrodermal(EDA) Response System}
    \label{fig:big_picture}
\end{figure}

However, challenges remain in generalizing and deepening EDA data analysis. The tonic component of EDA data ranges from 0.01$\mu$S to 40$\mu$S, and the phasic component from 0.01$\mu$S to 5$\mu$S, necessitating systems capable of detecting a wide conductivity range with high signal resolution in the same time\cite{sanchez2021correlation}. Due to user comfort and integration with devices like smartwatches, the placement of electrodes should be compromised to the wrist or main body rather than fingers and toes, result in higher demand for dynamic range and resolution. Some solutions involve high-resolution standalone Analog-to-Digital Converters (ADCs), leading to increased cost and power consumption, resulting in bulkier devices or reduced battery life \cite{garbarino2014empatica}. Others fail to achieve high-resolution EDA signals, compromising prediction accuracy \cite{udovivcic2017wearable}. This paper shows the design and demonstration of a high-accuracy, low-power electrodermal activity (EDA) signal acquisition system featuring an adaptive gain mechanism. The performance of the system was assessed using simulations and a custom-designed printed circuit board (PCB), achieving an error rate below 1\% and a Pearson correlation coefficient of 0.73 in comparison to the Empatica E4 \cite{garbarino2014empatica}. The system also demonstrated a power consumption of only 700$\mu$A under a 3.7V power supply.

\section{Working Principle} \label{Working Principle}
\begin{figure}[ht!]
    \centering
    \includegraphics[width= .95\linewidth,height=5cm]{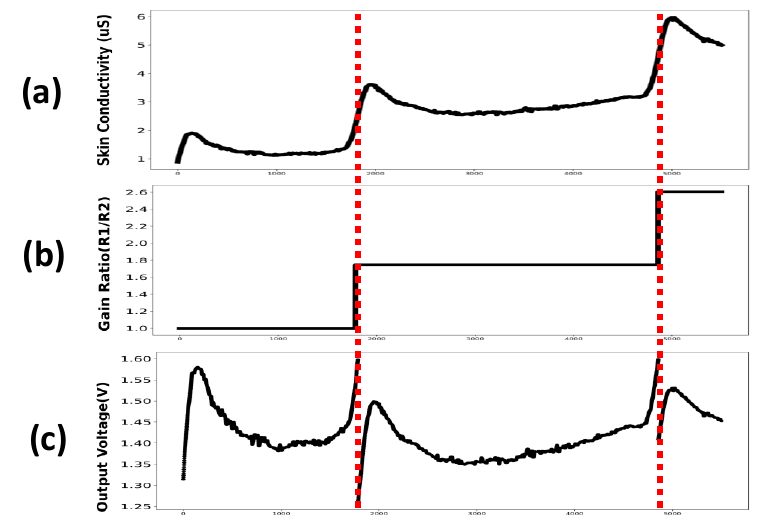}
    \caption{Simulated compensation mechanism illustration (a).skin conductivity (b).Gain Ratio (c).Output voltage}
    \label{fig:Simulated compensation mechanism illustration}
\end{figure}

As we mentioned before. EDA signals, spanning a range of 0.1 $\mu$S to 40 $\mu$S (resistance range from 10 M$\Omega$ to 25 K$\Omega$). Addressing this, we propose an EDA acquisition system with an adaptive gain selection mechanism using two multiplexers (MUXs) to select appropriate resistors (R1 and R2), connected to operational amplifier (OP) circuits. This setup enhances the system's dynamic range, with the Gain Ratio determined by the values of R1 and R2, inversely affecting the output voltage, as detailed in Section \ref{Discussion}. Figure. \ref{fig:Simulated compensation mechanism illustration} shows simulated EDA values from 1-6 $\mu$S using a Python library\cite{makowski2021neurokit2}, with fluctuations in skin conductivity affecting the output voltage. Upon saturation at 1.8V, the Gain Ratio adjusts from 1.0 to 1.8 to prevent overload, as depicted in Figure. \ref{fig:Simulated compensation mechanism illustration} (b). The system's firmware supports a sampling rate of 8 Hz (125 ms intervals) according to our previous research\cite{hosseini2022low, fang2023introducing}. During this period, the microcontroller (MCU) retrieves and processes the output voltage, calculates resistance values, transmits the data via Bluetooth, and adjusts R1 and R2 accordingly, to prepare for the next data cycle.

\section{Prototype Development} \label{Prototype Development}
In our system architecture, a customized PCB was developed, as illustrated in Fig. \ref{fig:Measurement Error}(b). At the core of our design is the nRF52832 MCU from Nordic, Inc., chosen for its multi-functionality, due to its integration of Bluetooth Low Energy (BLE) capabilities and an ARM Cortex-M4 processor with a floating-point unit, operating at a frequency of 64 MHz. The nRF52832's blend of processing power and wireless communication efficiency makes it particularly well-suited for the intricate requirements of our EDA acquisition system. For the power management system, the TP4054 chip (NanJing Top Power ASIC Corp) was implemented as the charging mechanism through a 5V power port. The system also incorporates two MAX4781 MUXs from Analog Devices, Inc., which are utilized to select the resistors that are connected to the operational amplifier (OP) circuits, specifically the TLV9061(TI, Inc). Two LM3671 DC-DC converter chips(TI, Inc), were employed as separate power sources for the MCU and the Analog Front End (AFE), both operating at 1.8V. This voltage level was specifically chosen to minimize power consumption across all components, aligning with our goal of achieving the lowest possible power usage.

\section{Accuracy and Power Consumption Evaluation on Board}

To evaluate system precision, we performed simulation-based resolution testing and on-board accuracy verification. Resolution is defined as ohms per bit, representing the resistor change corresponding to a one-bit ADC alteration. A lower value indicates better resolution and higher sensitivity to small resistor changes.

\begin{figure}[ht!]
    \centering
    \includegraphics[width= \linewidth, height = 8cm]{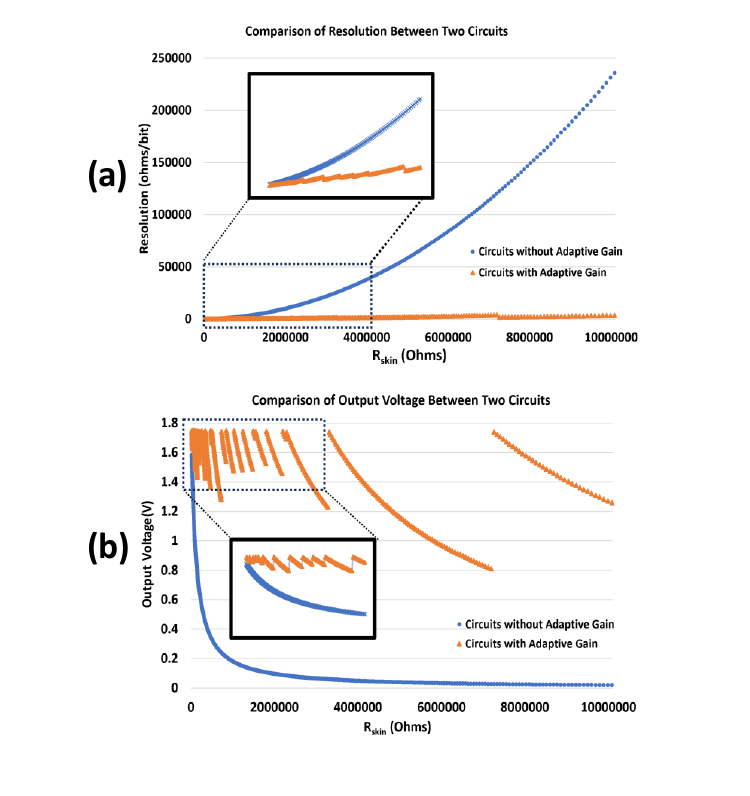}
    \caption{(a).Simulated Resolutions Between circuits with and without  Adaptive gain. (b). Comparison of Output Voltage Between circuits with and without Adaptive gain}
    \label{fig:Simulated Resolutions}
\end{figure}

Initially, we simulated an AFE configuration without adaptive gain, mapping skin resistance ranging from 25k$\Omega$ to 10M$\Omega$ to an output voltage between 0 and 1.8V (the full range of the ADC). This is illustrated by the blue line in Fig. \ref{fig:Simulated Resolutions}(b). Subsequently, we computed the resolution corresponding to each resistor value, as depicted by the blue line in Fig. \ref{fig:Simulated Resolutions}(a). The resolution begins at 34 $\Omega$/bit for a skin resistance of 25k$\Omega$ and escalates to 311 k$\Omega$/bit at 10M$\Omega$. On the other hand, the simulated results for our AFE with adaptive gain (shown as the orange line in Fig. \ref{fig:Simulated Resolutions}(a)) demonstrate significant improvements in resolution, particularly noticeable beyond the 300K$\Omega$ skin resistance threshold. This resolution starts at 31 $\Omega$/bit at 25k$\Omega$ and rise to a mere 4.6k$\Omega$/bit at 10M$\Omega$. A detailed comparison, particularly evident before 4M$\Omega$, is visible in the zoomed-in section of Fig. \ref{fig:Simulated Resolutions}(a). Here, the orange line appears in discontinuous segments, each representing different gain settings. Fig. \ref{fig:Simulated Resolutions}(b) compares the output voltages of the AFE with and without adaptive gain. Notably, the output voltage from the AFE with adaptive gain is discontinuous, reflecting the implementation of various gains. Each continuous segment within this graph correlates to a specific gain setting, underscoring the dynamic adaptability of our system. 

\begin{figure}[ht!]
    \centering
    \includegraphics[width=.8\linewidth]{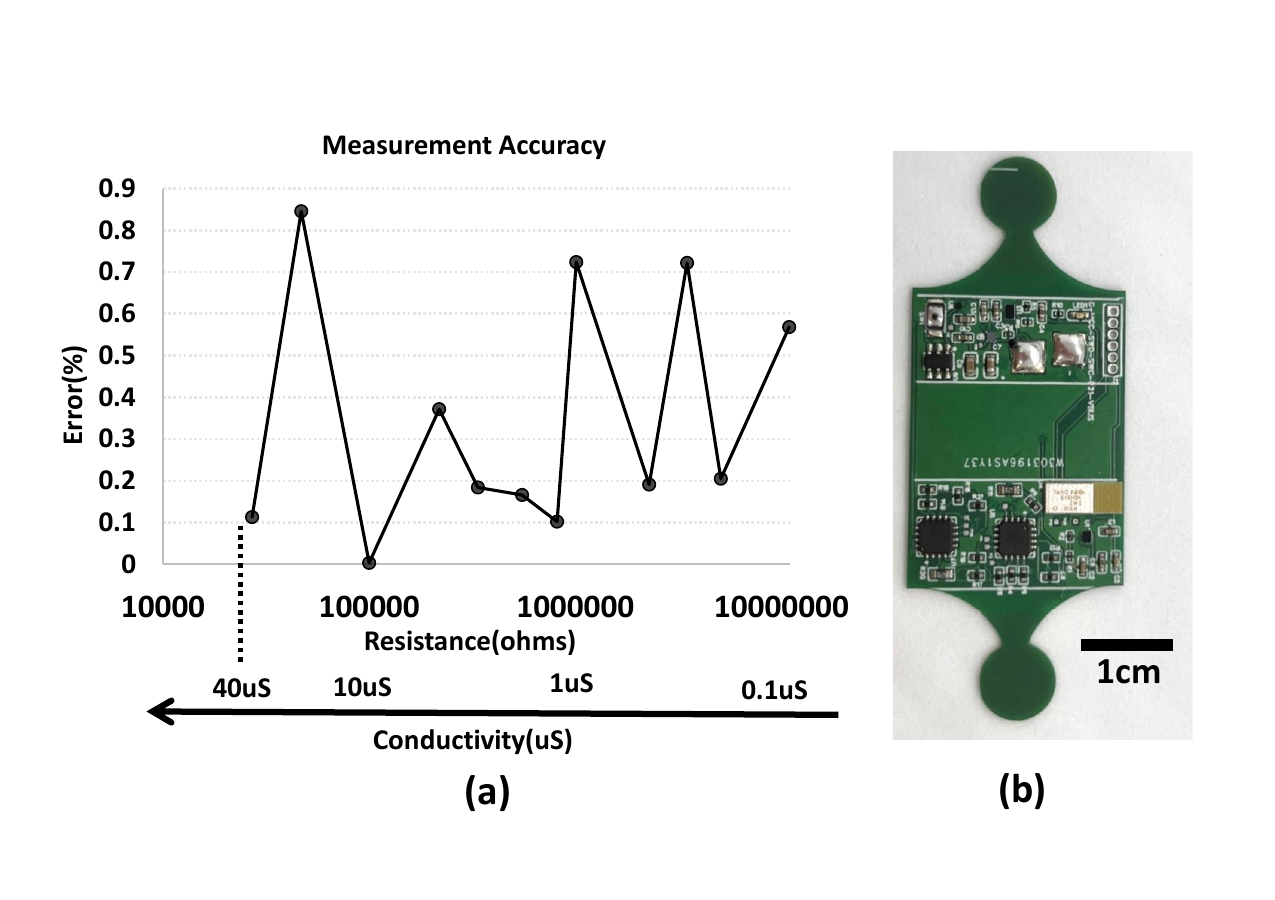}
    \caption{(a). Measurement Errors (b). Customized PCB for EDA Acquisition System}
    \label{fig:Measurement Error}
\end{figure}

\begin{figure}[ht!]
    \centering
    \includegraphics[width=\linewidth]{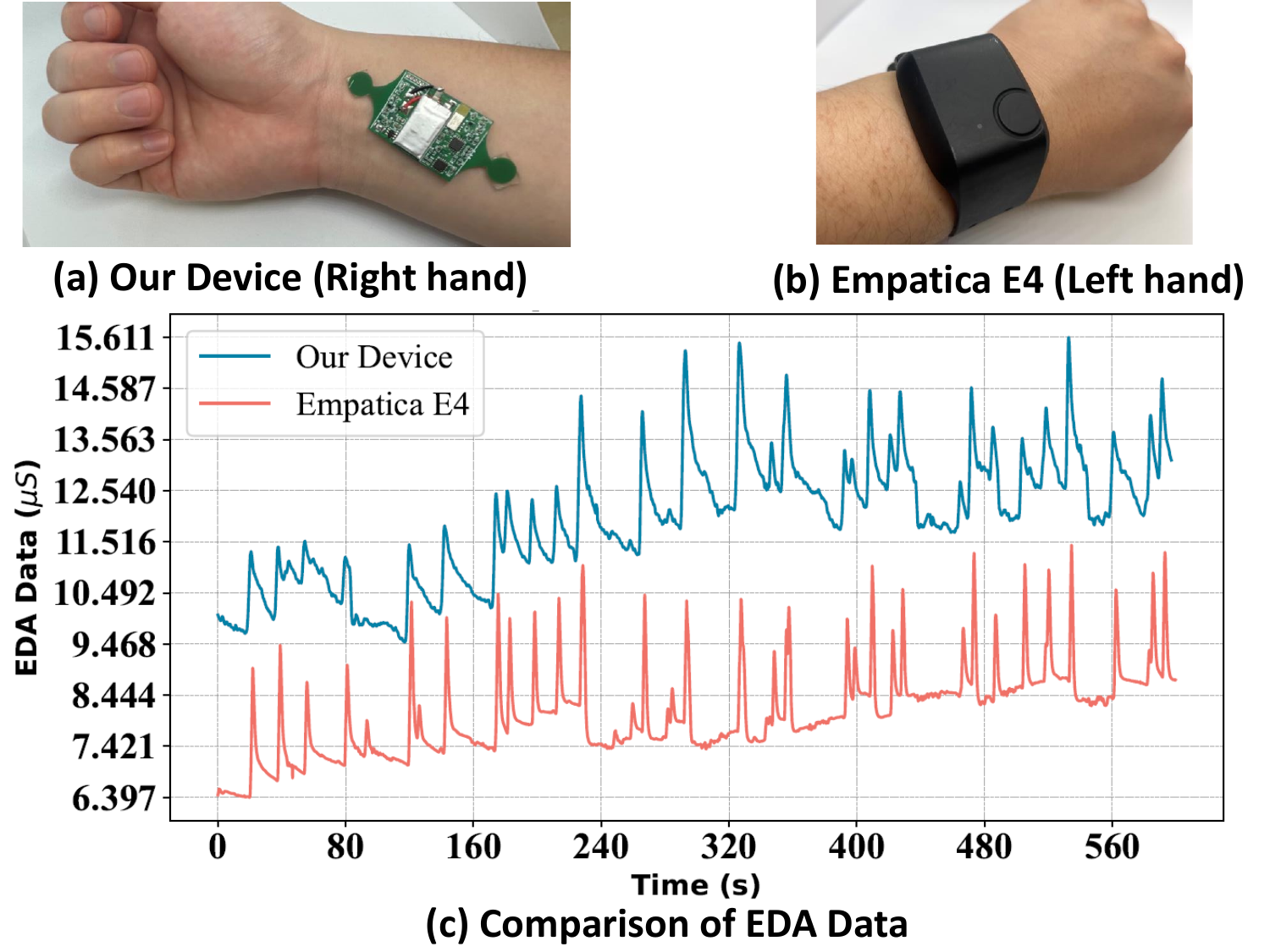}
    \caption{(a) Our device (b) Empatica E4. (c) Comparison of EDA Data}
    \label{fig:Comparison of EDA Data}
\end{figure}

Further, we conducted accuracy assessments of our system using a printed circuit board (PCB), as shown in Fig. \ref{fig:Measurement Error}(b). We utilized a variety of standard metal film resistors, specifically labeled as 27k, 47k, 100k, 220k, 330k, 560k, 820k, 1M, 2.2M, 3.3M, 5.1M, and 10M$\Omega$. These resistors were connected to the Analog Front End (AFE) to simulate different levels of skin resistance. The testing process involved the MCU selecting the appropriate resistors R1 and R2 based on the detected resistance levels, then calculating the resistor values from the ADC readings. To gauge the system's accuracy, we compared the true resistor values with those calculated from the ADC, analyzing the absolute differences as depicted in Fig. \ref{fig:Measurement Error}(a). The results show that error values across all tested resistors remained below 1\%. It's noteworthy that the absolute error values tend to increase with higher skin resistance values—a trend typical in such measurements, highlighting the need for precision in higher resistance ranges. Additionally, to evaluate the dynamic responsiveness of our system to Electrodermal Activity (EDA) signals, we conducted a comparative test against the Empatica E4 device \cite{mccarthy2016validation}. The Empatica E4 was worn on the left hand, as shown in Fig. \ref{fig:Comparison of EDA Data}(a), while our system was attached to the right hand, as depicted in Fig. \ref{fig:Comparison of EDA Data}(b). During a 10-minute test involving activities like deep breathing, breath holding, and speaking to induce EDA fluctuations, the results are presented in Fig. \ref{fig:Comparison of EDA Data}(c). These outcomes demonstrate our system's ability to effectively capture both peaks and trends within the EDA data. Moreover, a Pearson Correlation coefficient of 0.73, calculated from data collected simultaneously from both hands, confirms the robustness of our system in accurately capturing dynamic EDA signals in real-time \cite{milstein2020validating}. Through these comprehensive evaluations, we've been able to demonstrate the enhanced accuracy and resolution capabilities of our system equipped with adaptive gain control.

\begin{figure}[ht!]
    \centering
    \includegraphics[width=.9\linewidth, height = 6cm]{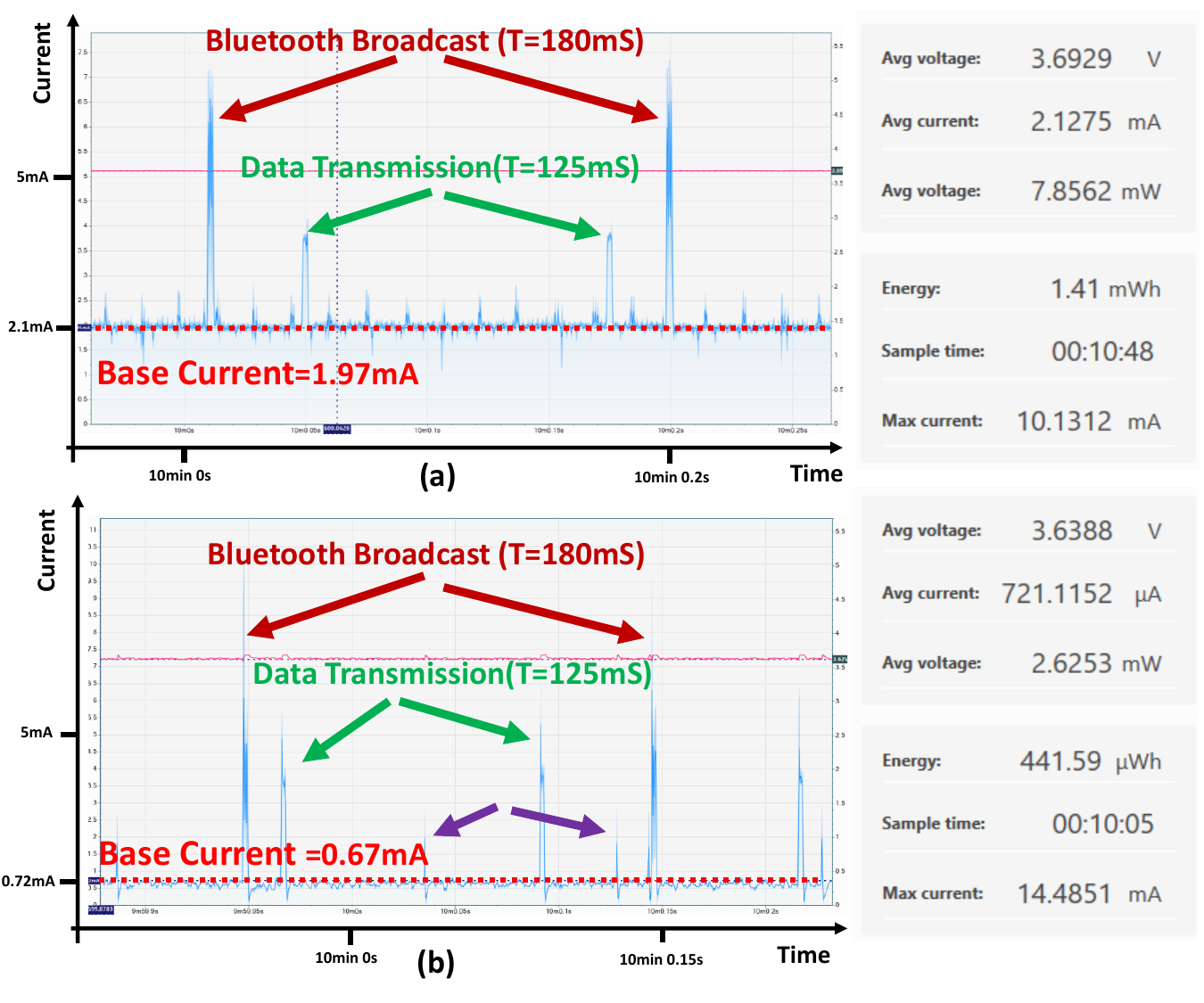}
    \caption{Power Consumption Analysis (a)Firmware with MUX, OP operating all time (b)Firmware with MUX, OP operating only as needed}
    \label{fig:Power Consumption Analysis}
\end{figure}

To optimize power consumption for long-term monitoring in wearable devices, we addressed the challenge from both hardware and firmware perspectives. In terms of hardware, we selected two DC-DC converters to efficiently reduce battery voltage from 3.5-4.2V to 1.8V, which are more efficient than low-dropout regulators (LDOs). Additionally, our design incorporates an enable function for the multiplexers (MUX) and operational amplifiers (OP), which are activated only during sampling periods and immediately disabled afterwards, significantly conserving energy. This strategy's effectiveness is illustrated in Fig. \ref{fig:Power Consumption Analysis}. The baseline scenario depicted in Fig. \ref{fig:Power Consumption Analysis}(a) shows the power usage with the MUX and OP continuously active, leading to notable current peaks due to Bluetooth activities and data transmissions, with a baseline current of 1.97mA and an average of 2.127mA over a 10-minute period. In contrast, Fig. \ref{fig:Power Consumption Analysis}(b) shows reduced power usage when the MUX and OP are only activated as needed, cutting the baseline current to 670$\mu$A and the average current to 721$\mu$A, confirming that the main power consumption originates from the sensing components rather than the MCU. Furthermore, the viability of the system for long-term use in compact wearable devices is supported by a 30-hour continuous data acquisition test using a single 30mAh Li-ion and depicted in Fig. \ref{fig:Comparison of EDA Data}(a). This approach not only ensures energy efficiency use but also maintains the essential accuracy and dynamic range.

\section{Discussion} \label{Discussion}
In the discussion section, we focus on the details of our proposed system's design, emphasizing the accuracy of skin resistance measurements and power efficiency. First, the tolerance of the gain selection resistors (R1, R2) is $\pm$1\%, leading to inaccuracies. For instance, a resistor labeled as 330K$\Omega$ (R1) measured 339K$\Omega$. To improve accuracy, each resistor was measured with an LCR meter, allowing the MCU to use these actual values for calculations instead of the labeled ones. Future designs might use high-precision thin-film chip resistors, like the P0402 series ($\pm$0.01\%), to avoid the need for individual calibration. Regarding power optimization, our initial design included two DC-DC converters. One was always active to power the MCU, and the other, with an enable function, powered the sensing system. We explored enabling and disabling the converter for the sensing system but observed an average current over 3.6mA, double that of the always-on scenario. This phenomenon is similar to an ignition effect in vehicles, where frequently starting the engine consumes more fuel than leaving it running, suggesting that re-enabling the converter uses more power than maintaining it in an always-on state. Future versions may use a single DC-DC converter with disabled MUX and OP to reduce power consumption. For Bluetooth communication, we implemented a custom protocol to consolidate 15 seconds of data into one package, transmitted every 15 seconds to save power. However, this caused occasional Bluetooth disconnections. Further testing showed that the power savings from this method were minimal, about 70$\mu$A, just a 10\% reduction. This suggests that extended intervals for data transmission do not significantly improve power efficiency as expected. These findings will inform future design improvements.

\section{Conclusion and Future Work}

In this paper, we developed an Electrodermal Activity (EDA) acquisition system designed for enhanced wearability. By focusing on precise accuracy and power optimization, the system selectively activates components during data collection to reduce energy consumption. These efforts were validated through rigorous long-term endurance tests, demonstrating the system’s suitability for extended use in compact wearable devices. This study contributes to the ongoing development of biometric monitoring, offering insights for future enhancements in wearable health technologies. Our system has the potential to be integrated into various forms, like flexible rubber bands or smart clothing, as suggested by \cite{goncu2021calmwear, zhang2024introducing,goncu2020therapeutic} and other compact wearable device.

\bibliographystyle{ieeetr}
\bibliography{EMBC}

\end{document}